%% file: main.tex
\documentclass[sigconf]{acmart}

\usepackage{dblfloatfix}
\usepackage{graphicx}
\usepackage{booktabs}
\usepackage{lipsum}
\usepackage{xcolor}
\usepackage{tabularx}
\usepackage[capitalise]{cleveref}

\usepackage{multicol}

\usepackage{array}

\AtBeginDocument{%
  }

\setcopyright{acmlicensed}
\copyrightyear{2024}
\acmYear{2024}
\acmDOI{10.1145/3640457.3688014}

\acmConference[ACM RecSys '24]{October 14--18,
  2024}{Bari, Italy}

\begin{document}

\title{Explainable Multi-Stakeholder Job Recommender Systems}

\author{Roan Schellingerhout} \orcid{0000-0002-7388-309X}
\affiliation{%
  \institution{Maastricht University}
  \streetaddress{Paul-henri Spaaklaan 1}
  \city{Maastricht}
  \state{Limburg}
  \country{The Netherlands}
  \postcode{6229 EN}
}
\email{roan.schellingerhout@maastrichtuniversity.nl}

\renewcommand{\shortauthors}{Roan Schellingerhout}

\begin{abstract}
Public opinion on recommender systems has become increasingly wary in recent years. In line with this trend, lawmakers have also started to become more critical of such systems, resulting in the introduction of new laws focusing on aspects such as privacy, fairness, and explainability for recommender systems and AI at large. These concepts are especially crucial in high-risk domains such as recruitment. In recruitment specifically, decisions carry substantial weight, as the outcomes can significantly impact individuals' careers and companies' success. Additionally, there is a need for a multi-stakeholder approach, as these systems are used by job seekers, recruiters, and companies simultaneously, each with its own requirements and expectations. In this paper, I summarize my current research on the topic of explainable, multi-stakeholder job recommender systems and set out a number of future research directions. 
\end{abstract}



\maketitle

\input{source/1_background}
\input{source/2_completed_research}
\input{source/3_future_work}

\begin{acks}
I would like to thank my supervisors, Francesco Barile and Nava Tintarev, for their guidance and support during my PhD research.
\end{acks}


\end{document}

%% file: source/1_background.tex
\section{Background and Context}
With recommender systems being one of the most widespread forms of machine learning used, they tend to be under heavy scrutiny by the public \cite{ricci2011introduction}. These systems are extensively utilized across various domains such as e-commerce, social media, and content streaming, making their impact on daily life significant. Consequently, concerns about privacy, bias, and transparency have become more pronounced \cite{pu2012evaluating}. Oftentimes, recommender systems are even distrusted, with users and representatives being wary of potential manipulation being performed by the system to nudge them into certain beliefs or behaviors \cite{stray2024building}. 


One way to address such suspicions is through the use of explainable artificial intelligence (XAI). By allowing users (and lawmakers alike) to gain insights into how specific recommendations came to be, we can enable them to understand the system better, leading to more trust in its efficacy and less suspicion of foul play \cite{arrieta2020explainable}. Explainable AI can be critical for gaining user trust, as well as compliance with regulations such as the GDPR and the EU AI Act \cite{EU2021ai, EuropeanParliament2016a, guidotti2018survey}.

While considerable research has been done on using explainable AI to aid system developers and users, further prominent stakeholders (e.g., advertising companies, item providers, lawmakers) should not be ignored. This need for a multi-stakeholder approach requires a nuanced approach, as it makes recommending and explaining items more complex \cite{abdollahpouri2019multi, abdollahpouri2020multistakeholder, abdollahpouri2021multistakeholder, bauer2019leveraging}. All stakeholders want recommendations and explanations optimized for their needs, but they often have conflicting interests. This balancing act becomes even more complex in high-risk domains, such as recruitment. Job recommender systems (JRSs), which match job seekers with potential employment opportunities, can have a considerable impact on individuals' lives \cite{de2021job}. Job recommender systems generally have three main stakeholders, all of which fall under the three main recommender system stakeholder types identified by \citet{abdollahpouri2020multistakeholder}: \textit{candidates} - the people looking for a job (i.e., consumers); \textit{companies} - businesses offering positions of employment (i.e., providers); and \textit{recruiters} - people whose job it is to match candidates and vacancies (i.e., system). Each of these stakeholders has different needs and priorities, making a multi-stakeholder approach to generating explanations crucial \cite{abdollahpouri2020multistakeholder, de2021job}. For example, candidates need to trust the system and understand why a job is suitable for them before making such an impactful decision \cite{schellingerhout2023co}. Proper explanations can also help mitigate biases and ensure fair treatment of all candidates, as it enables the system to be scrutinized. Furthermore, recruiters can use explanations to understand why certain candidates are recommended, allowing them to focus more efficiently on promising matches. Companies, on the other hand, can be enabled to quickly find the most relevant candidate from a large pool of options, increasing their productivity. 

This leads us to formulate the following research question for my PhD project: \textit{How can we create an explainable, multi-stakeholder job recommender system that supports targeted explanations for different stakeholders?}

To assist in answering this research question, we consider the following sub-questions:
\begin{description}
    \item[SQ1:] What are the stakeholder-specific explanations requirements and preferences of candidates, recruiters, and companies respectively?
    \item[SQ2:] How can we design an explainable multi-stakeholder job recommender system that outperforms state-of-the-art systems in user- and provider-side performance metrics?
    \item[SQ3:] To what extent does the inclusion of explainability into a real-world job recommender system improve its perceived usefulness, transparency, and trust?
\end{description}

In the rest of this paper, we first summarize the research conducted so far. Then, we set out multiple directions for future work. 

%% file: source/2_completed_research.tex
\section{Completed research}

The research conducted during the PhD project up until this point has focused on SQ1 (finding stakeholder requirements and preferences) and SQ2 (building an explainable multi-stakeholder job recommender system). In this section, we describe the specific experiments we have performed so far.  

\subsection{Stakeholder preferences (SQ1)}
To get an initial indication of the explanation preferences of the three main stakeholder types, we conducted a small-scale user study wherein we interviewed 6 participants while we exposed them to different examples of possible explanation types \cite{schellingerhout2023co}. We used this user study as a starting point to get an indication of what explanation types were most promising to explore in more depth in future work. While we used a relatively small sample size, this allowed us to spend a considerable amount of time with each participant (around 1 hour per person), which enabled us to have the participants co-design the explanation types. 

We found considerable preference differences, both \emph{between} and \emph{within} stakeholder types. Candidates and recruiters strongly preferred textual explanations over visual explanations, mentioning that those were easier to grasp and had a more `personal' feel to them. Company representatives initially gravitated towards the textual explanations too, as those were easiest to understand at first. However, they indicated a preference towards visual, graph-based explanations once they had spent some time trying to grasp those. Once they understood how the graph-based explanations should be interpreted, they mentioned how such visualizations allowed them to get a comprehensive overview of the complex relations in the data at a glance. This difference between stakeholders could largely be attributed to the fact that company representatives tended to have more experience with working with charts and graphs as part of their day-to-day job. 

However, there was significant disagreement between members of each stakeholder type as well. For example, recruiters disagreed on how comprehensive the explanations should be - either preferring long texts allowing them to provide sufficient detail to their clients when trying to convince them of potential matches, or preferring more limited explanations to offer them an initial indication of suitability, after which they could use their expertise to come to a more honed-in decision. To allow users to cater the explanation environment to their personal preferences and needs, we determined that interactive interfaces are crucial for job recommendation, as those allow individuals to access the data they find important, while not getting overwhelmed by information they do not consider useful.

\subsection{Mock-up System Experiment (SQ1)}
\begin{figure*}
    \centering
    \includegraphics[width=\textwidth]{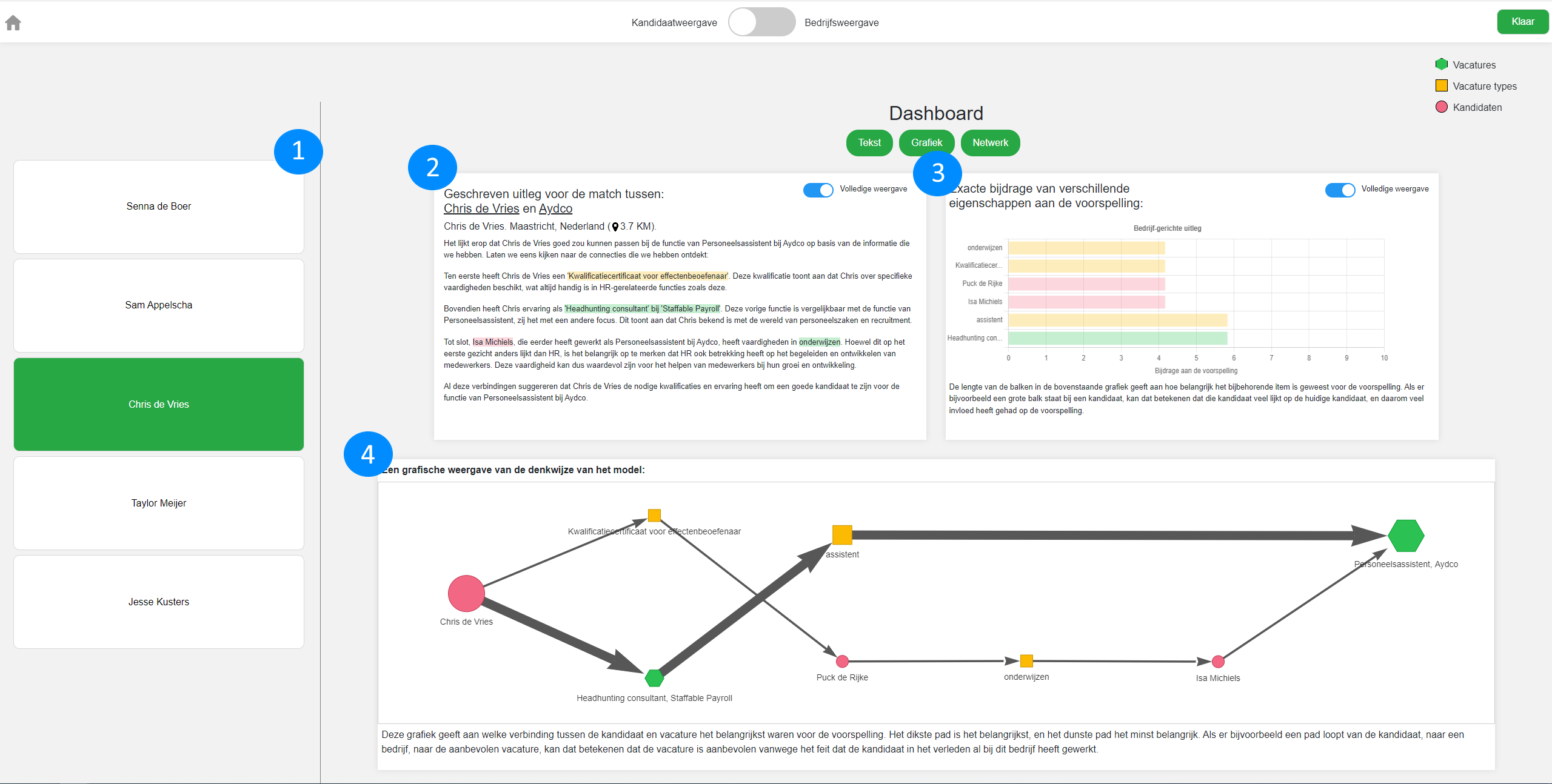}
    \caption{The interface of the online environment with which the participants interacted. In this screenshot, all explanations are enabled. These can individually be toggled based on the user's preference. The web environment uses exclusively Dutch text, as the interviewees were all native Dutch speakers. The environment consists of the following components: (1) the list of recommended items, which were presented in a randomized order (i.e., the top item was not necessarily the best match); (2) the textual explanation; (3) the bar chart explanation; (4) the graph-based explanation. This example shows a \textit{real} explanation.}
    \label{fig:interface}
\end{figure*}

After having co-designed the different explanation types, we created a prototypical explanation environment wherein users could browse multiple recommended items and their accompanying explanations (\cref{fig:interface}). We tested this explanation environment with 30 participants in total; 10 of each stakeholder type. When interacting with the environment, both subjective (perceived usefulness, trust, and transparency) and objective (correctness and efficiency) metrics were collected. Participants were tasked to select what they considered to be the best option from the list of items twice - once after having seen real explanations generated by a graph neural network, and once after having seen random explanations (they were shown the different explanations in random order). Due to the nature of the data available to us at the time, this system did not allow users to get recommendations for their personal CV or vacancy, but rather had them read a pre-selected CV or vacancy before seeing the recommended items, after which they were instructed to decide as if they were the person/company whose CV/vacancy they just read.

We found that preferences largely stayed the same, with candidates and recruiters strongly gravitating towards the textual explanations, and company representatives having a more diverse range of preferences. However, regardless of which explanation type the participants preferred, the difference in metrics between the random and real explanations was very limited. I.e., whether a user was shown a `nonsensical' random explanation or a genuine explanation, their opinion of the system barely changed. While the subjective metrics trended upward with the real explanations, this trend was not statistically significant. Correctness, on the other hand, even went \emph{down} when participants were using real explanations to come to a decision. One contributing factor to this lack of difference is that the most commonly preferred explanation type, text, lends itself quite poorly to indicating minor differences between examples (i.e., it is hard for users to spot discrepancies in phrasing between explanations, e.g., `somewhat' instead of `strongly'). Furthermore, the decrease in correctness indicates that users do not actively engage with the explanations and instead apply their own reasoning to the situation. Even if the system gives a specific argument for why an item is a good match, users often come up with widely different reasons for their decision, even when agreeing with the model. This lack of engagement leads to a slight benefit for the random explanations, as those are less likely to create `friction', allowing participants to always apply their own reasoning to their decision without feeling like they disagree with the model. 

Based on these findings, we determine that we should instead provide \emph{decision-support} to the users of a job recommender system. While most XAI research focuses on persuasive explanations, we find that trying to persuade the stakeholders of the model's correctness is futile, as they will apply their domain expertise to the decision-making process regardless. As such, it is better to support them in this process, rather than trying to steer them in a certain direction. This also addresses another concern in job recommendation: ground truth values are often generated manually by human recruiters - as a result, they are not \textit{objective} truths. Attempting to force users to agree with the system can therefore be counterproductive, as their decision could, in theory, be preferable over the one determined to be `correct'. Furthermore, during the experiment, multiple participants indicated wanting a clear, direct relation between the explanation and the `source material' (i.e., the CV or vacancy). If the explanation contained information that was not (directly) present in the CV/vacancy (e.g., work experience stored in the data, but not written in the CV), participants tended to get confused, wondering where that new information came from. Therefore, we conclude that the arguments used by the decision-support system should be clearly grounded in the source material. 

\subsection{Explainable Graph Neural Network (SQ2)}
The aforementioned papers made use of a rather low-quality, but publicly available dataset. As a result, the performance of the recommender system, and as a consequence its explanations, had significant room for improvement. To address this shortcoming, we collaborated with a large, international recruitment agency in order to gain access to a high-quality, manually labeled, proprietary job recommendation dataset. To determine the efficacy of graph neural networks (GNN) on this dataset, we implemented a novel explainable GNN, the \textit{Occupational Knowledge-based Recommender using Attention} (OKRA). We then compared OKRA to multiple state-of-the-art job recommendation models; both text-only and other graph-based models. 

Our experiments showed that graph-based models strongly outperformed text-only models. Considering the majority of JRS research focuses primarily on text-based solutions, this finding could have considerable consequences on the field. While most research focuses on utilizing transformer-based models to compare CV and vacancy texts to find matches, our findings indicate that this leaves a significant amount of predictive power unused. Compared to state-of-the-art graph-based models, OKRA performed significantly better due to its ability to make stakeholder-specific decisions (as candidates and companies are not necessarily always in agreement), but at the cost of increased training time. Additionally, we found that most state-of-the-art models are slightly biased against both rural candidates and companies, indicating a need for the consideration of regional fairness in the field of job recommendation. 

While OKRA is inherently explainable, the focus of this paper was on its recommendation performance rather than its explainability. Due to the architecture we used, OKRA is able to generate multiple explanations for a single prediction, meaning it can separate `positive' and `negative' contributions to a decision. While this theoretically lends OKRA's explanations to a decision support system, we did not evaluate the model's explainability component, leaving it for future work. 

%% file: source/3_future_work.tex
\section{Future work}
During the rest of the project, we will primarily focus on improving the model and its explanations so that they conform to the stakeholders' demands as much as possible. In the rest of this section, we set out multiple avenues for future research related to each sub-question. 

\subsection{SQ1: Designing desirable explanations}
Based on the research we have conducted for sub-question 1 so far, we already have a general understanding of the preferences and needs of the different stakeholders. However, we have also identified multiple shortcomings with our previous approach that need to be addressed. To conclusively answer SQ1, we will focus on alleviating these shortcomings in future work, so that we can present a concrete implementation that adheres to all stakeholder requirements. 

\subsubsection{Improving explanation coherence}
One of the main difficulties faced by the stakeholders in our mock-up experiment was that they struggled to connect the explanations to the source material. Considering the explanations were generated using both the source material \emph{and} structured data, parts of the explanation based on the structured data did not necessarily align with what was shown in the CV or vacancy (e.g., it highlighted a skill that someone did not list on their CV). This led to confusion and made it more difficult for participants to understand the explanations. To address this shortcoming, we will attempt to integrate the CVs/vacancies and the structured data so that their contents are more aligned. 

One possible approach is to make use of automated knowledge graph construction from text \cite{bosselut2019comet, wang2021data, wu2019automatic}. For this, we would create a machine learning pipeline capable of automatically finding entities and their underlying relations in a given text. The graphs generated from this text could then be linked to the rest of the structured data, so that for every piece of structured data, there exists a link to a word or phrase in the source material. This would not only improve the coherence of the textual explanations but also allow the model to more directly integrate the information stored in the CV/vacancy into the recommendation process. However, this approach would require some type of training data, as zero-shot learning is likely to be insufficiently integrated into the existing ontology. Alternatively, we could apply untrained clustering algorithms to cluster the embeddings of different tokens, so that different tokens referring to the same concept can be coalesced. While this approach does not require training data, it is presumed to be less reliable, as mismatches between the structured and unstructured data could still occur. When only using textual data to generate the knowledge graph, it is certain that all the information in the graph is also stored in the text, however, when combining structured and unstructured data, even with clustering, some tokens/concepts come up in one data type, but not the other (e.g., work experience that is stored in the structured data, but not mentioned in the CV). 
 
\subsubsection{Clarifying textual explanations}
Furthermore, we found that substantive differences in attention weights can lead to rather minute differences in the textual explanations. As mentioned above, this made it difficult for users to differentiate between textual explanations with different contents correctly. While an attention weight of 0.7 instead of 0.2 stands out immediately, properly communicating this difference without referring to the exact values (since referring to numeric values directly was indicated as complicated and overwhelming by stakeholders) can be difficult for LLMs. For example, while describing these values as `moderately high' and `fairly low' is correct, such formulations do not stand out immediately, which causes users to easily gloss over them. To solve this, we intend to fine-tune an LLM, such as GPT-4, on a collection of explanations that have been manually verified as `clear' or `understandable'. To determine what constitutes a clear explanation, we will conduct an experiment wherein participants will be asked to pick a preferred option between two versions of the same explanation, but with the value of one textual feature (such as word count, word complexity, level of formality, etc.) altered. By repeating this match-up multiple times, each time with different features being changed, we can finally determine a user's preferred explanation type (e.g., high word count, low word complexity, low formality, etc.) Given a sufficient sample, we can then determine what type of explanation is preferred by the end users.

\subsection{SQ2: Improving model performance}
\subsubsection{Exploiting linked data}
One major benefit of using knowledge graphs is that they are capable of easily combining data from multiple sources \cite{bizer2011linked}. While regular databases can be difficult to combine, primarily due to differences in (naming) conventions and higher-order relations being hard to implement using relational algebra, knowledge graphs easily allow data from multiple sources to be combined. One major aspect of job recommendation where this can make a large difference, is in the initial creation of node embeddings. Currently, all node embeddings used by OKRA were initialized randomly, except for those based on CVs or vacancies. The CV and vacancy nodes had a starting embedding based on the text embedding value created by a transformer-based model. By incorporating linked data, node types that currently do not have any text related to them, such as function titles and codes (e.g., \textit{international standard classification of occupation},\footnote{\url{https://isco.ilo.org/en/}} or ISCO, codes), can be linked to their respective nodes in existing graphs like that of WikiData\footnote{\url{https://www.wikidata.org/}} and DBpedia.\footnote{\url{https://www.dbpedia.org/}} These linked data sources often have extensive descriptions of the functions/codes, allowing the model to use those descriptions to create starting embeddings. Furthermore, these data sources contain significantly more data than most domain-specific datasets, using which the knowledge graph can be made more exhaustive, enabling more high-level relations in the data to be identified. 

\subsection{SQ3: Evaluating the system as a whole}
\subsubsection{Evaluating in a real-world context}
The explanation evaluations we have done so far have all been conducted using non-personal data. The CVs and vacancies used in our mock-up system were examples using which the participants had to role-play. While this was sufficient for the scope of this study, using such data can lead to some bias, as it is more difficult to make a decision on behalf of someone else than for oneself. Logically, we aim to address this in future work by creating a live version of the environment, wherein users will be able to submit their own vacancy/CV and personal data, so that the model can generate a personalized list of recommended items for them. This will enable the users to go through a more natural decision-making process, as they do not have to bear the additional cognitive load of having to remember a CV/vacancy that is not theirs. As a result, we will be able to evaluate the system in a more holistic manner using this live version of the environment; users can interact with the system as they would in a real-world scenario as well, making it possible for us to determine to what extent the users interact with the explanations (rather than simply reading the recommended vacancies/CVs like they would with a non-explainable environment). Furthermore, by having a live, working system, we will be able to more easily experiment with a large sample size, as participants will be able to interact with the system independently on their own time (which was not possible with the mock-up version, as there they needed to be supervised). 

We will again need to collaborate with a recruitment agency to conduct such a live evaluation, which may cause difficulties (or even be impossible within our time span). This leads to another avenue of future research: determining to what extent our findings generalize to other domains. Although our project focuses specifically on JRSs, many overlaps exist between the recruitment domain and domains such as dating. While the stakeholders' exact requirements are likely to differ, we do expect general similarities to exist between the two groups. As a result, it would be interesting to additionally evaluate our findings in a different domain that is similar in nature, to assess the generalizability of our findings.

\subsubsection{List-wise explanations}
Lastly, a significant challenge we anticipate when shifting towards decision-support-focused explanations is having to present explanations related to a list of recommendations. When users are shown the pros and cons of different recommended items, understanding why one item is ranked higher than another can be complex. While providing pair-wise comparisons is relatively straightforward, offering clear and comprehensive explanations in a list-wise context is much more challenging \cite{heuss2024rankingshap}. This difficulty arises from the need to show the intricate relationships and trade-offs among multiple items simultaneously. To address this issue, we will explore ways to effectively communicate the advantages and disadvantages of multiple items in future research. This could involve developing new comparative visualization techniques, interactive interfaces, or summary metrics that can help users grasp the overall ranking rationale and make informed decisions based on the recommendations provided.